\documentclass[%
 reprint,
 superscriptaddress,
 amsmath,amssymb,
 aps,
]{revtex4-2}

\usepackage{graphicx}
\usepackage{dcolumn}
\usepackage{bm}
\usepackage{hyperref}
\usepackage{cleveref}
\usepackage{enumerate}
\usepackage{float}

\usepackage{subcaption}

\usepackage[mathlines]{lineno}

\usepackage{custom}
\usepackage{xcolor}
\definecolor{lilac}{RGB}{199, 164, 219}

\begin{document}


\title{
Deep-learning classification of physically admissible nuclear-matter equations of state}

\author{Ahmed Abuali}\email{amabuali@uh.edu | ahmed.sabri.abuali@gmail.com}
\affiliation{
 Department of Physics, University of Houston, Houston, TX 77204, USA
}

\author{Micheal Kahangirwe}\email{mkahangi@kent.edu | kahangirwem@gmail.com}
\affiliation{
Center for Nuclear Research, Department of Physics, Kent State University, Kent, OH 44243, USA
}

\author{Francesco Di Clemente}
\affiliation{
Department of Physics, University of Houston, Houston, TX 77204, USA
}

\author{Vianney E Diaz-Barraza}\affiliation{Department of Electrical \& Computer Engineering, The University of Texas at El Paso, TX 79968, USA}

\author{Jorge A Munoz}\affiliation{Department of Physics, The University of Texas at El Paso, TX 79968, USA}

\author{Claudia Ratti}
\affiliation{
 Department of Physics, University of Houston, Houston, TX 77204, USA
}

\date{\today}

\begin{abstract}
Thermodynamic stability and causality impose fundamental physics constraints on the equation of state (EoS) of nuclear matter. Conventionally, verifying these constraints requires explicitly evaluating thermodynamic observables such as the specific heat, baryon-number susceptibility, and speed of sound, which can become computationally expensive when large numbers of candidate EoSs are explored. In this work, we investigate whether the normalized pressure surface, $Q(T,\mu_B)={P(T,\mu_B)}/{T^4}$, contains sufficient information to determine the physical admissibility of an EoS without explicitly evaluating these quantities. We develop a supervised convolutional neural network (CNN) that receives only this pressure representation as input and classifies EoSs as physically admissible or inadmissible. The network is trained using labels obtained from direct thermodynamic stability and causality checks, but it is not provided with any information about the internal parameters of the theoretical framework used to generate the EoSs. We first train and evaluate the model using EoSs generated within an Ising-mapping framework, achieving an overall classification accuracy of $97.65\%$ on previously unseen test data. We then repeat the procedure using EoSs generated within a distinct holographic framework and obtain a perfect classification of the corresponding independent test set. These results demonstrate that the pressure surface alone contains characteristic geometric signatures of thermodynamic stability and causality violations that can be learned directly by a CNN. Because the proposed classifier relies exclusively on the pressure surface, it is largely independent of the underlying EoS-generation framework and avoids the need to evaluate higher-order thermodynamic observables during inference. We further show that the machine-learning validation pipeline is approximately 20 times faster than the direct validation when the normalized pressure surface is supplied directly as a two-dimensional array. Therefore, our results establish a fast approach for identifying physically admissible equations of state that does not require knowledge of the internal parameters of the underlying EoS-generating framework, and demonstrate that the information encoded in the pressure surface is sufficient to diagnose thermodynamic consistency.

\end{abstract}

\maketitle

\section{Introduction}
The equation of state (EoS) of strongly interacting matter plays a central role in understanding the thermodynamic properties of Quantum Chromodynamics (QCD). It characterizes the relationship among pressure, temperature, and conserved charge chemical potentials, thereby determining the macroscopic behavior of nuclear matter. The EoS is essential both for interpreting experimental observables from relativistic heavy-ion collisions \cite{HADES:2019auv,BRAHMS:2004adc,STAR:2005gfr,PHOBOS:2004zne,PHENIX:2004vcz,Jahan:2026hvs} and for modeling dense matter in compact astrophysical objects such as neutron stars and neutron star mergers \cite{Most:2022wgo,ReinkePelicer:2025vuh}. In hydrodynamic simulations of heavy-ion collisions, the EoS describes the evolution of the hot medium created in the collision, while multimessenger observations of neutron stars and gravitational waves provide complementary constraints on the properties of matter at high baryon density. 

At sufficiently high temperatures and/or baryon densities, nuclear matter undergoes a transition from hadronic degrees of freedom to a deconfined quark–gluon plasma (QGP). Lattice QCD calculations at vanishing baryon chemical potential have shown that this transition is a smooth crossover~\cite{Aoki:2006we}. At finite baryon density, however, first-principles calculations remain challenging due to the sign problem \cite{Philipsen:2012nu}. Consequently, theoretical studies rely on effective models and extrapolation techniques to explore the QCD phase diagram in the full temperature-baryon chemical potential ($T$, $\mu_B$) plane \cite{MUSES:2023hyz,Sorensen:2023zkk,ReinkePelicer:2025vuh,Jahan:2026hvs}. Many of these approaches predict a richer phase structure, including the possible existence of a critical point separating a crossover region from a first-order phase transition line. The search for signatures of such critical behavior remains one of the primary objectives of the Beam Energy Scan (BES) program at the Relativistic Heavy Ion Collider (RHIC), which has recently completed its data-taking phase and is now focused on data analysis. Future experimental facilities are expected to further explore this region of the QCD phase diagram~\cite{Bzdak:2019pkr,Pandav:2022xxx}.

From a theoretical perspective, a wide variety of frameworks have been developed to construct the QCD EoS, based on available constraints such as those listed in Ref. \cite{MUSES:2023hyz}. These include effective models of QCD thermodynamics such as holographic approaches based on the gauge/gravity duality~\citep{Gubser:2008yx,DeWolfe:2010he,DeWolfe:2011ts,Hippert:2023bel,Critelli:2017oub,Grefa:2021qvt, Rougemont:2023gfz,Yang:2026brr}, functional methods such as Dyson–Schwinger or functional renormalization group techniques~\citep{Fu:2019hdw,Gunkel:2021oya,Gao:2020fbl}, and lattice-based extrapolation schemes that extend first-principle results to finite baryon density~\citep{Basar:2023nkp,Shah:2024img,Borsanyi:2021sxv,Shah:2026pue,Abuali:2025tbd, Allton:2003vx,Allton:2005gk,Borsanyi:2012cr,Bazavov:2017dus,Bollweg:2022fqq}. Many of these constructions are inherently parametric, introducing a number of free parameters that control the thermodynamic behavior of the system and the structure of the phase diagram. While this flexibility allows one to explore a broad range of physical scenarios, not all parameter choices lead to physically admissible EoSs. In particular, pathological regions may emerge in which thermodynamic stability and/or causality are violated, especially in models that incorporate critical phenomena. Such EoSs cannot be reliably used in phenomenological or astrophysical simulations.

Physical consistency of an EoS is governed by thermodynamic stability and causality requirements. The stability conditions of the EoS can be reduced to two conditions, positive specific heat $C_V \geq 0$ and positive baryon number susceptibility $\chi_2^B \geq 0$, which ensure convexity of the relevant thermodynamic potentials. Causality further requires that the speed of sound remains subluminal, $0 \leq c_s^2 \leq 1$. These conditions are determined by derivatives of the pressure with respect to temperature and chemical potential. In practice, verifying these conditions requires explicitly computing higher-order derivatives of the pressure surface $P(T,\mu_B)$ on discretized grids. For complex parametric constructions, especially when scanning large multidimensional parameter spaces, this procedure can become computationally expensive and sensitive to numerical noise.

Recent advances in machine learning (ML), particularly deep learning methods, have provided powerful tools for identifying patterns and extracting structures from high-dimensional datasets in many areas of physics. In statistical and many-body physics, deep learning approaches have been successfully applied to classify phases of matter and detect phase transitions directly from raw data without explicit knowledge of the order parameters \cite{Abuali:2025vmn, Abuali:2026lqm,Carrasquilla:2016oun,Wang:2016nmn,vanNieuwenburg:2016zsd,Li:2017xaz,Hu:2017wey,Wetzel:2017olt,Rodriguez-Nieva:2018cbl,Foreman:2018ktj,Cossu:2018pxj,ShibaFunai:2018aaw,Tan:2019eih,Bachtis:2020dmf,Walker:2020hiq,Han:2022mhy,Li:2025rbf}. Similar techniques have found increasing applications in high-energy and nuclear physics, including studies of lattice QCD \cite{Boyda:2020hsi,Kanwar:2020xzo} and jet substructure \cite{Baldi:2014kfa, Kasieczka:2019dbj} . These developments suggest that ML methods may offer efficient alternatives to traditional diagnostic approaches by learning directly from physical observables, thereby avoiding the need for explicit derivative-based calculations.

Recent work has begun to explore ML-based approaches for identifying physically consistent EoSs. For example, Ref.~\cite{Mroczek:2022oga} employed active learning to classify unstable or acausal parameter choices in the lattice-based Taylor expansion known as the BEST EoS~\cite{Parotto:2018pwx}, which incorporates an embedded 3D Ising critical point through a linear mapping. While such approaches can efficiently identify valid regions of parameter space, they remain tied to specific model parameterizations and therefore depend on prior knowledge of the internal variables used to construct the EoS. As a result, their applicability is inherently limited to the particular theoretical framework used to generate the training data, making generalization across different EoS constructions difficult.

In this work, we pursue a complementary and more general strategy. Rather than learning the relationship between internal model parameters and physical admissibility within a specific EoS construction, we investigate whether the physical admissibility of an EoS can be inferred directly from the normalized pressure surface,
$Q(T,\mu_B)=P(T,\mu_B)/T^4$. The pressure surface is a physical observable common to all EoS constructions and provides a framework-independent representation that does not depend on the internal parameters used to generate the EoS. We therefore ask whether a machine-learning model trained solely on this information can learn the geometric signatures associated with thermodynamic consistency and distinguish physically admissible from inadmissible EoSs.

From a thermodynamic perspective, stability and causality constraints are ultimately encoded in the geometric structure of the pressure surface. Consequently, violations of these constraints may leave identifiable signatures in the pressure surface itself. If such signatures can be extracted directly from $Q(T,\mu_B)$, the resulting classifier would provide a framework-independent alternative to approaches that operate in model-parameter space and could significantly reduce the computational cost of EoS validation.

To investigate this question, we develop a supervised convolutional neural network (CNN) classifier that receives as input only the normalized pressure surface, 
$Q(T,\mu_B)=P(T,\mu_B)/T^4$. Training labels are obtained from explicit thermodynamic stability and causality checks, allowing the network to learn directly from examples of physically admissible and inadmissible EoSs without explicitly providing derivative-based thermodynamic information. Because the classifier operates solely on the pressure surface, it does not require access to the internal parameters used to generate the EoS, and can therefore be applied independently of the underlying theoretical framework. 

We evaluate the approach using two distinct frameworks of EoS constructions. In the first, the network is trained following standard machine-learning train-validation-test procedures using EoSs generated within an Ising-based framework, and is subsequently subjected to an additional independent testing stage using previously unseen EoSs from the same framework. In the second, the same training, validation, testing, and independent testing procedure is repeated using EoSs generated within a holographic framework \cite{Hippert:2023bel}, merged to a Hadron Resonance Gas model EoS according to Ref. \cite{Yang:2026brr}. In both cases, the classifier achieves excellent performance on independently generated EoSs that were not used during training, validation, testing, or model selection, demonstrating that this representation contains sufficient geometric information to identify physically admissible EoSs. We further show that the trained classifier provides a computationally efficient alternative to direct stability and causality checks, reducing the cost of EoS validation while remaining independent of the underlying EoS-generation framework and its internal parameterization.

The remainder of this paper is organized as follows. In Sec.~\ref{sec:Physical Constraints}, we review the thermodynamic stability and causality conditions used to determine the physical admissibility of the EoS. Section~\ref{sec:EoS} describes the EoS datasets employed in this work, including the Ising-based and holographic constructions used throughout model development and independent testing. Section~\ref{sec:Methodology} presents the machine-learning methodology, including data representation and preprocessing, neural-network architecture, and the training strategy. In Sec.~\ref{sec:Results}, we report the classification performance obtained for both EoS frameworks and examine the ability of the model to identify physically admissible EoSs directly from the normalized pressure surface. We also compare the computational cost of the machine-learning validation pipeline with conventional validation based on explicit thermodynamic stability and causality checks. Finally, Sec.~VI summarizes our conclusions and discusses possible future directions.

\section{Physical Constraints on the Equation of State}
\label{sec:Physical Constraints}

Before introducing the neural network framework, we briefly review the physical requirements for a physically admissible EoS. These criteria are used in this work to assign physical admissibility labels and to distinguish physically consistent behavior from pathological behavior.

\subsection{Thermodynamic Stability}

Thermodynamic stability requires the thermodynamic potential to be convex with respect to its natural variables. When expressed in terms of the pressure $P(T,\mu_B)$, this requirement translates into positivity conditions on the second derivatives of the pressure with respect to temperature and baryon chemical potential.

The stability properties of the system can be examined through the Hessian matrix of second derivatives of the pressure,
\begin{equation}
H =
\begin{bmatrix}
\frac{\partial^2 P}{\partial T^2} & \frac{\partial^2 P}{\partial \mu_B \partial T} \\
\frac{\partial^2 P}{\partial T \partial \mu_B} & \frac{\partial^2 P}{\partial \mu_B^2}
\end{bmatrix}
=
\begin{bmatrix}
\frac{\partial s}{\partial T} & \frac{\partial n_B}{\partial T} \\
\frac{\partial n_B}{\partial T} & \chi_2^B
\end{bmatrix},
\end{equation}
where $s$ is the entropy density, $n_B$ is the baryon number density, and $\chi_2^B$ denotes the baryon number susceptibility. 

Thermodynamic stability requires this matrix to be positive definite. In practice, these conditions reduce to requiring positive specific heat at constant volume, $C_V$ and positive baryon number susceptibility $\chi_2^B$,

\begin{align}
C_V &= \frac{T}{\chi_2^B} \left[
\left( \frac{\partial s}{\partial T} \right)_{\mu_B} \chi_2^B 
- \left( \frac{\partial n_B}{\partial T} \right)_{\mu_B}^2
\right] > 0, \\
\chi_2^B &= \left( \frac{\partial^2 P}{\partial \mu_B^2} \right)_T > 0.
\end{align}

Violating any of these conditions corresponds to unstable thermodynamic behavior, such as negative compressibility or runaway energy fluctuations, and therefore indicates that the corresponding EoS cannot represent a physically realizable equilibrium system. Since these conditions are determined by derivatives of the pressure surface, such violations are expected to leave characteristic local geometric distortions in the pressure field itself.

\subsection{Causality}

In addition to thermodynamic stability, relativistic systems must satisfy causality. In hydrodynamic descriptions, this requirement is expressed through the speed of sound $c_s^2$,
\begin{equation}
c_s^2 = \left(\frac{\partial P}{\partial \epsilon}\right)_{s/n_B},
\end{equation}
which characterizes the propagation speed of small pressure perturbations in the medium.

Relativistic causality requires that the speed of sound remains subluminal,
\begin{equation}
0 \le c_s^2 \le 1,
\end{equation}
where natural units ($c=1$) are used. Values of $c_s^2 > 1$ correspond to superluminal propagation of signals and therefore violate causality.

When expressed in terms of derivatives of the pressure with respect to $T$ and $\mu_B$ along $s/n_B$ trajectory necessary for heavy-ion physics, the speed of sound can be written as \cite{Floerchinger:2015efa}
\begin{eqnarray}
c_s^2(T,\mu_B) &&
= \left(\frac{\partial P}{\partial \epsilon}\right)_{s/n_B} \nonumber \\
&&= \frac{n_B^2\frac{\partial^2 P}{\partial T^2}
- 2 s n_B \frac{\partial^2 P}{\partial T\partial \mu_B}
+ s^2 \frac{\partial^2 P}{\partial \mu_B^2}}
{(\epsilon + P)\left(
\frac{\partial^2 P}{\partial T^2}\frac{\partial^2 P}{\partial \mu_B^2}
-\left(\frac{\partial^2 P}{\partial T\partial \mu_B}\right)^2
\right)}.
\end{eqnarray}

Therefore, evaluating causality also requires knowledge of second derivatives of the pressure surface. As in the case of thermodynamic stability, violations of these causality conditions are expected to alter the local geometric structure of the pressure field. Together, these stability and causality conditions provide the criteria used throughout this work to assign admissibility labels to the EoS datasets employed for model development and independent testing.

\section{Equation of State Input}
\label{sec:EoS}

A variety of EoSs have been proposed in the literature~\cite{ReinkePelicer:2025vuh,Jahan:2026hvs}. In this Section, we briefly describe the two frameworks of generating EoSs used in this study. The first is an Ising-based construction developed within the MUSES collaboration that incorporates critical behavior through a mapping of the three-dimensional Ising universality class onto QCD thermodynamics \cite{Kahangirwe:2024cny}. The second is a holographic framework merged with a hadron resonance gas (HRG) description \cite{Yang:2026brr}. Together, these datasets provide physically distinct EoS constructions and enable us to assess whether physical admissibility can be inferred directly from the pressure surface, independent of the underlying EoS-generation framework.

\subsection{Ising 2D-TExS}

The Ising 2D-TExS framework provides a lattice-informed description of strongly interacting matter that explicitly incorporates a QCD critical point, making it a realistic and nontrivial testbed for our study. It extends the thermodynamic coverage into the finite baryon density region by combining a lattice-constrained $T'$-expansion with universal critical scaling mapped from the 3D Ising model.

A defining feature of this framework is the implementation of a tunable QCD critical point belonging to the 3D Ising universality class. Near criticality, the divergence of the correlation length implies that macroscopic behavior is governed by global symmetries rather than microscopic details. Consequently, systems sharing the same symmetry structure exhibit identical critical scaling behavior. This universality principle was systematically exploited in~\cite{Kahangirwe:2024cny} to map the critical behavior of the 3D Ising model onto the QCD phase diagram via a non-universal transformation,
\begin{figure*}[!thb]
    \centering
    \includegraphics[width=\textwidth]{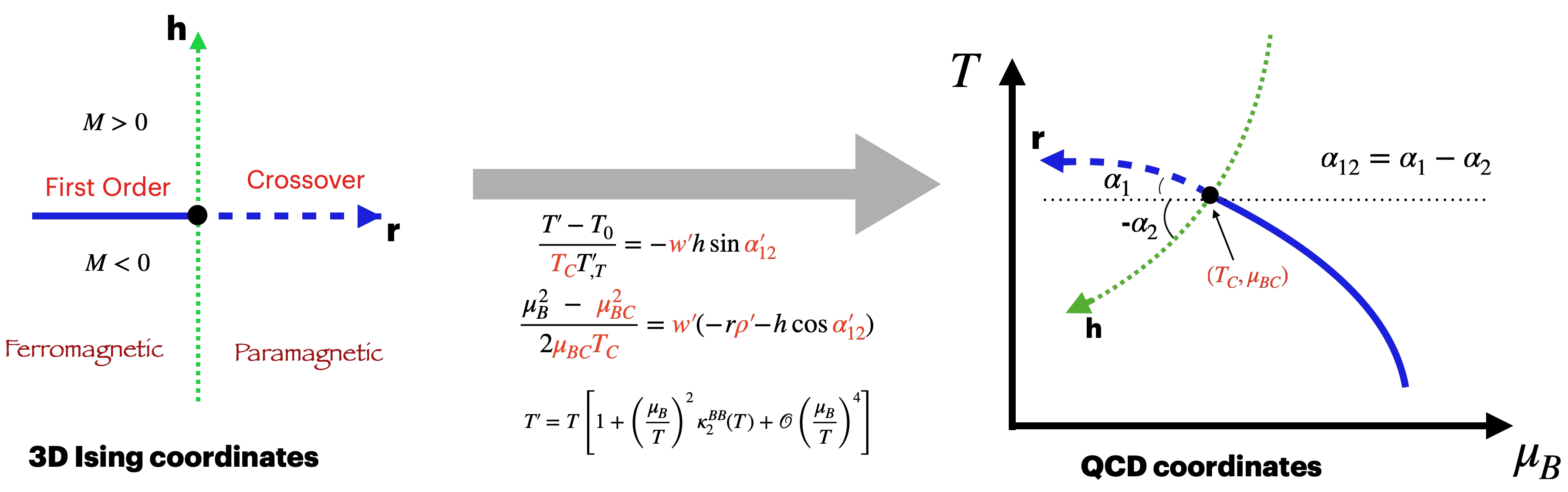}
    \caption{
    Left: 3D Ising model axes with a critical point at $(r=0,h=0)$. Right: corresponding QCD coordinates with a critical point located at $(T_C,\mu_{BC})$. The parameters $\mu_{BC}$, $w'$, $\rho'$, and $\alpha'_{12}$ control the position, orientation, and scaling of the mapping (Figure from Ref. \cite{Kahangirwe:2024cny}).
    }
    \label{fig:mapping}
\end{figure*}
where $(r,h)$ denote the reduced temperature and magnetic field of the Ising model. The mapping parameters $\mu_{B_C}$, $w'$, $\rho'$, and $\alpha'_{12}$ control the location, orientation, and scaling of the critical region in QCD coordinates, as illustrated in Fig.~\ref{fig:mapping} and are directly related to $\mu_{B_C}$,$w$, $\rho$, $\alpha_{12}$ as in~\cite{Parotto:2018pwx} for linear mapping.

Although these parameters determine the properties of the critical region within the Ising-TExS framework, they are not provided to the neural network and are used only in the generation of the EoS dataset.

While lattice QCD provides first-principles results at vanishing baryon chemical potential, its direct extension to finite density is limited. The Ising-TExS framework overcomes this limitation by combining the universal critical construction with an improved extrapolation of lattice data using the $T'$-expansion scheme~\cite{Borsanyi:2021sxv}. In this approach, the baryon density is written as
\begin{equation}
    n_B(T,\mu_B) = \hat{\mu}_B \, \chi_2^B\!\left(T'(T,\mu_B)\right),
    \label{Eqn:Tprime_nB}
\end{equation}
where $\hat{\mu}_B = \mu_B/T$ and the effective temperature $T'$ is expanded as
\begin{equation}
T'(T,\mu_B) = T\left(1 + \kappa_2 \hat{\mu}_B^2 + \kappa_4 \hat{\mu}_B^4 + \cdots\right).
\end{equation}

This redefinition resums higher-order density contributions into a modified temperature argument, enabling a controlled and stable extrapolation beyond the radius of convergence of conventional Taylor expansions. It preserves lattice constraints at $\mu_B=0$ while extending the EoS into the finite-density regime relevant for phenomenology.

A key observation is that $\chi_2^B(T,0)$ is analytic across the crossover, implying that any non-analytic behavior associated with a critical point must enter through $T'(T,\mu_B)$. Accordingly, the effective temperature is decomposed as
\begin{equation}
T' = T'_{\rm reg} + T'_{\rm crit},
\end{equation}
where $T'_{\rm reg}$ encodes the smooth lattice background and $T'_{\rm crit}$ captures the singular critical contribution \cite{Kahangirwe:2024cny}. The latter is constructed to reproduce the universal scaling behavior of the 3D Ising model, while the regular component is fixed such that the full $T'$ matches lattice QCD constraints at $\mu_B=0$.

To consistently embed this structure into QCD coordinates, a two-step mapping procedure is employed. First, the mapping is constructed in the $(T,\mu_B)$ space such that the first-order phase transition line is aligned along a fixed reference temperature. Second, the lattice-based temperature $T'_{\text{lat}}$ replaces $T$, inducing the physically required curvature of the critical line in QCD coordinates, as illustrated in Fig.~\ref{fig:mapping}.

Substituting the full $T'(T,\mu_B)$ into Eq.~\eqref{Eqn:Tprime_nB} ensures that the baryon density inherits the correct critical scaling behavior in the vicinity of the QCD critical point, while remaining anchored to lattice QCD results along the $\mu_B = 0$ axis.

The pressure is then obtained by integrating in $\mu_B$ at fixed temperature,
\begin{equation}
  \hat{P}(T,\mu_B)
  =
  \hat{P}_{\rm lat}(T,0)
  +
  \int_{0}^{\mu_B} d\hat{\mu}_B'\;
  \hat{n}_B(T,\mu_B')\,,
\end{equation}
where $\hat{P}_{\rm lat}(T,0)$ fixes the integration constant and guarantees thermodynamic consistency with lattice results at vanishing baryon density.

All remaining thermodynamic observables follow from derivatives of $P(T,\mu_B)$ with respect to $T$ and $\mu_B$, including higher-order
baryon number susceptibilities such as $\chi_2^B$. These observables
are subsequently used to evaluate the thermodynamic stability and
causality conditions discussed in Sec.~II and to assign physically admissible or inadmissible labels to the generated EoS's.

By construction, the Ising-TExS EoS consistently unifies lattice-constrained thermodynamics, controlled finite-density extrapolation, and symmetry-driven critical scaling within a single framework. This makes it a robust and physically well-motivated tool for studying QCD matter at finite temperature and baryon chemical potential.

\subsection{Other Frameworks}

Pathological EoSs are not limited to a single theoretical construction. For example, in holographic approaches, which employ gauge/gravity duality to study QCD-like systems, certain parameter choices may lead to unstable or acausal behavior \citep{Hippert:2023bel}. Similarly, functional renormalization group methods \citep{Fu:2019hdw} and other effective descriptions may produce EoSs with unphysical properties under specific conditions. Thus, there is a need for general methods capable of assessing the physical admissibility of EoSs across different theoretical frameworks.

In addition to the Ising-TExS dataset described above, we therefore consider a second dataset generated within a holographic framework merged with a hadron resonance gas (HRG) description \cite{Yang:2026brr}. This framework differs substantially from the Ising-TExS construction in both its physical motivation and mathematical formulation, with free parameters motivated in this case by the merging procedure. The resulting EoSs are labeled using the same thermodynamic stability and causality conditions introduced in Sec.~\ref{sec:Physical Constraints}. By performing separate classification studies on both the Ising-based and holographic datasets using the same input representation, neural-network architecture, and training procedure, we can assess whether physical admissibility can be inferred directly from the temperature-normalized pressure surface, independent of the underlying EoS-generation framework.

\section{Methodology}
\label{sec:Methodology}

We formulate the problem as a binary classification task in which each EoS is labeled as either physically admissible or inadmissible according to the thermodynamic stability and causality conditions described in Sec.~\ref{sec:Physical Constraints}. The neural network is not given access to the internal parameters of the model used to generate the EoS, nor is it provided with explicitly computed thermodynamic derivatives. Instead, each EoS is represented only by the normalized pressure surface

\begin{equation}
Q(T,\mu_B)=\frac{P(T,\mu_B)}{T^4}
\end{equation}

defined on the two-dimensional $(T,\mu_B)$ grid. This choice is motivated by two considerations. First, \(Q=P/T^4\) is a standard dimensionless representation of the pressure in finite-temperature QCD thermodynamics, where the dominant thermal scaling of the pressure is approximately proportional to \(T^4\). Second, dividing by \(T^4\) reduces the large numerical hierarchy present in the raw pressure surface and makes local geometric distortions more visible to the convolutional network. The central question is therefore whether the normalized pressure surface \(Q(T,\mu_B)\), by itself, contains sufficient information for identifying the signatures of thermodynamic instability and causality violation.

For each EoS, the input to the neural network is a single-channel array containing \(Q(T,\mu_B)\). The labels are obtained from direct thermodynamic checks of \(C_V\), \(\chi_2^B\), and \(c_s^2\), but these quantities are used only for labeling and are not provided to the model during inference.

\subsection{Data representation and preprocessing}

Each EoS is stored as a two-dimensional numerical array defined on a discretized $(T,\mu_B)$ grid. The pressure values are first converted into their normalized representation, producing a single-channel input array for the convolutional neural network.

To ensure numerical stability during training, a second normalization step is applied using global statistics computed exclusively from the training subset. Specifically, the mean $\bar{Q}$ and standard deviation $\sigma_Q$ are calculated over the training data, and each sample is standardized according to

\begin{equation}
\tilde{Q}(T,\mu_B)=\frac{Q(T,\mu_B)-\bar{Q}}{\sigma_Q}.
\end{equation}

\subsection{Neural-network architecture}

\begin{figure*}[!t]
\centering
\includegraphics[width=\textwidth]{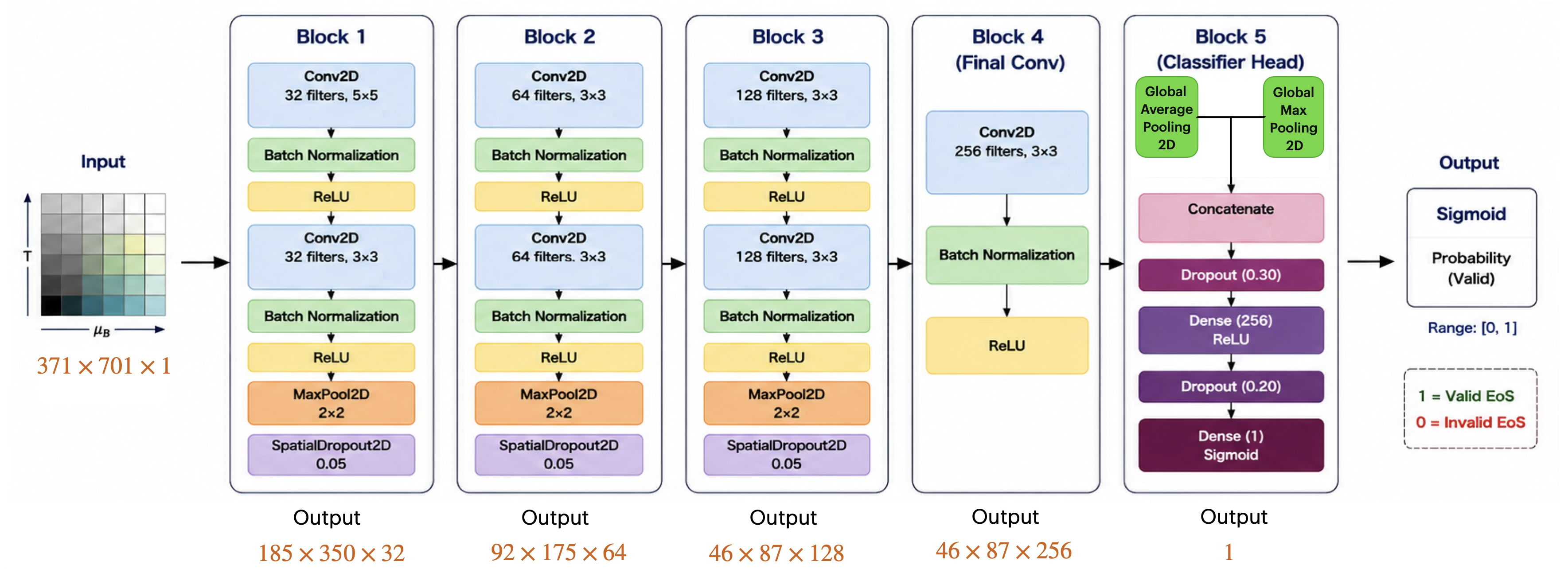}
\caption{
Architecture of the convolutional neural network used for binary classification of equation-of-state (EoS) surfaces. The input is a single-channel tensor containing the normalized pressure field $Q(T,\mu_B)=P(T,\mu_B)/T^4$ defined on the two-dimensional $(T,\mu_B)$ grid. The network consists of four convolutional blocks with progressively increasing feature depth, followed by parallel global average pooling and global max pooling operations whose outputs are concatenated and passed through a fully connected classifier. The final sigmoid output represents the probability that a given EoS is physically admissible.
}
\label{fig:Architecture}
\end{figure*}

A schematic representation of the convolutional neural network architecture used in this work is shown in Figure~\ref{fig:Architecture}.

The network receives as input a single-channel tensor of shape $(371,701,1)$ corresponding to the standardized normalized pressure field defined on the $(30~\text{MeV}\leq T \leq 400 ~\text{MeV},0\leq \mu_B\leq 700 ~\text{MeV})$ grid of stepsize 1 MeV for both $T$ and $\mu_B$. As illustrated in Figure~\ref{fig:Architecture}, the architecture consists of four sequential convolutional blocks with progressively increasing numbers of filters. In the first three blocks, each stage contains two convolutional layers, each followed by batch normalization and a ReLU activation function. Max-pooling operations are then applied to progressively reduce the spatial dimensionality while preserving the dominant local structures, and spatial dropout regularization is included to improve generalization and reduce overfitting. The final convolutional block increases the feature depth without additional pooling, allowing the network to construct higher-level representations from the extracted local patterns.

Following the convolutional feature extraction stage, the final feature maps are processed through two complementary global pooling operations: global average pooling and global max pooling. These parallel operations allow the network to capture both distributed geometric information and highly localized features that may be associated with thermodynamic instabilities or causality violations. The resulting pooled features are concatenated and passed through a dense layer with ReLU activation, with dropout regularization applied before and after the dense layer, followed by the final sigmoid output neuron, which returns the probability that a given EoS belongs to the physically admissible class.

The hierarchical structure of the network allows early convolutional layers to identify local geometric distortions in the pressure surface, while deeper layers progressively combine these features into higher-level representations associated with patterns relevant to thermodynamic stability and causality. This design enables the classifier to learn directly from the geometry of the pressure surface itself, without requiring prior knowledge of the explicit thermodynamic conditions used to assign the labels.

\subsection{Training procedure}

The neural network is trained as a supervised binary classifier using labeled datasets consisting of physically admissible (Valid) and inadmissible (Invalid) EoSs. Separate training procedures are carried out independently for each physical framework considered in this work, allowing the classifier to be evaluated within each framework individually.

In addition to the standard training and validation procedure, an additional independent testing stage is performed after model development is completed. In this stage, the finalized model is evaluated on entirely separate EoS datasets that are not used during training, validation, hyperparameter optimization, or threshold selection. The purpose of this independent testing stage is to assess the ability of the classifier to generalize beyond the specific samples used during model development and to verify that the learned decision boundaries remain robust when applied to newly generated EoSs drawn from the same physical framework.

To examine the data efficiency and robustness of the learning process, each framework is studied using two independent training configurations with different dataset sizes. For the Ising-based framework, the model is trained using both a larger dataset and a substantially reduced dataset, while an analogous two-stage training procedure is performed independently for the holographic framework. This allows us to evaluate the sensitivity of the learned classifier to reductions in the number of available training samples.

Training is performed by minimizing the binary cross-entropy loss function using an Adam-based optimization algorithm. To improve generalization and account for possible numerical uncertainties in the labeling procedure, a small amount of label smoothing $\epsilon=0.01$  is applied during training. Since the validity labels are determined through numerical evaluations of thermodynamic conditions involving higher-order derivatives, small numerical uncertainties may arise near the boundary separating valid and invalid regions. The label smoothing procedure helps reduce sensitivity to such numerical uncertainties during training.

To reduce overfitting and improve convergence stability, early stopping is employed so that training is automatically terminated once validation performance ceases to improve. In addition, the learning rate is adaptively reduced during later stages of training to allow finer optimization near convergence.

The network produces a continuous output probability ($p_{valid}\in[0,1]$), representing the predicted likelihood that a given EoS satisfies the physical admissibility conditions. Final binary classification is obtained using a decision threshold selected on the validation set by maximizing the balanced accuracy between the valid and invalid classes.
All machine-learning models were implemented in Python using the TensorFlow/Keras framework. The machine-learning code used in this work is publicly available in the Nuclear-EoS-Validity-CNN GitHub repository~\cite{Abuali_Nuclear_EoS_Validity_CNN}. Model training was performed on NVIDIA GPU hardware.

\vspace{-1.5em}

\section{Results and discussion}
\label{sec:Results}

In this Section, we evaluate the performance of the convolutional neural network in identifying physically admissible EoSs from the normalized pressure surface \(Q(T,\mu_B)\). We first compare the classification performance obtained within the Ising-based and holographic frameworks using both the full and reduced training datasets. We then examine the effect of training-set size on the independent testing performance and discuss the differences in model performance observed between the two physical frameworks. For the Ising-based model, we additionally present classification maps in the \((w,\rho)\) parameter space. Finally, we compare the computational cost of direct physics-based validation with ML-based validation using pressure inputs at different stages of preprocessing.

\subsection{Classification performance across physical frameworks}

We first examine the ability of the neural network to distinguish between physically admissible (Valid) and inadmissible (Invalid) EoSs within each physical framework independently. Two separate training configurations are considered for each framework in order to evaluate the classification performance under different amounts of available training data.

For the Ising-based framework, the model is trained using two different datasets containing either 4000 or 1000 EoSs per class. The corresponding classification results obtained on the validation dataset are summarized in Table~\ref{tab:validation_summary}, while the independent testing results on previously unseen Ising-generated equations of state are reported in Table~\ref{tab:independent_summary}. In both cases, the network achieves high classification accuracy, with overall independent testing accuracies of 97.65\% and 95.95\%, respectively.

For the holographic framework, analogous experiments are performed using training datasets containing either 2000 or 500 EoSs per class. The corresponding validation and independent testing results are also summarized in Tables~\ref{tab:validation_summary} and~\ref{tab:independent_summary}. Remarkably, the network achieves perfect classification performance in both training configurations, yielding near-perfect separation between the two classes, with 100\% classification accuracy observed for both validation and independent testing datasets.

These results demonstrate that the normalized pressure surface alone contains sufficient information for highly accurate identification of physical admissibility conditions. The consistently strong agreement between validation and independent testing performance further indicates that the learned classifier captures robust geometric signatures associated with thermodynamic stability and causality rather than overfitting to specific training samples.

For each independently trained model, the final binary decision threshold is determined from the corresponding validation dataset by scanning a range of threshold values and selecting the value that maximizes balanced classification accuracy. The resulting threshold is then fixed and used for the subsequent independent-testing analysis.

\begin{widetext}

\begin{center}

\refstepcounter{table}\label{tab:validation_summary}

\textbf{TABLE~\Roman{table}:} Validation-set classification performance for both frameworks with balanced classes. Precision, recall, and F1-score are macro-averaged.

\vspace{0.5em}

\begin{tabular}{lccccc}
\hline
Framework & Training samples/class & Precision & Recall & F1-score & Accuracy \\
\hline
Ising-based & 4000 & 0.9758 & 0.9756 & 0.9756 & 0.9756 \\
Ising-based & 1000 & 0.9554 & 0.9550 & 0.9550 & 0.9550 \\
Holography-based & 2000 & 1.0000 & 1.0000 & 1.0000 & 1.0000 \\
Holography-based & 500 & 1.0000 & 1.0000 & 1.0000 & 1.0000 \\
\hline
\end{tabular}

\end{center}

\vspace{1.2em}

\begin{center}

\refstepcounter{table}\label{tab:independent_summary}

\textbf{TABLE~\Roman{table}:} Classification performance on independent test datasets consisting of previously unseen equations of state excluded entirely from model development.

\vspace{0.5em}

\begin{tabular}{lcccc}
\hline
Framework & Training samples/class & Valid-class accuracy (\%) & Invalid-class accuracy (\%) & Overall accuracy (\%) \\
\hline
Ising-based & 4000 & 98.45 & 96.85 & 97.65 \\
Ising-based & 1000 & 97.80 & 94.10 & 95.95 \\
Holography-based & 2000 & 100.00 & 100.00 & 100.00 \\
Holography-based & 500 & 100.00 & 100.00 & 100.00 \\
\hline
\end{tabular}

\end{center}

\end{widetext}

\begin{figure*}[t]
    \centering
    \includegraphics[width=0.48\textwidth]{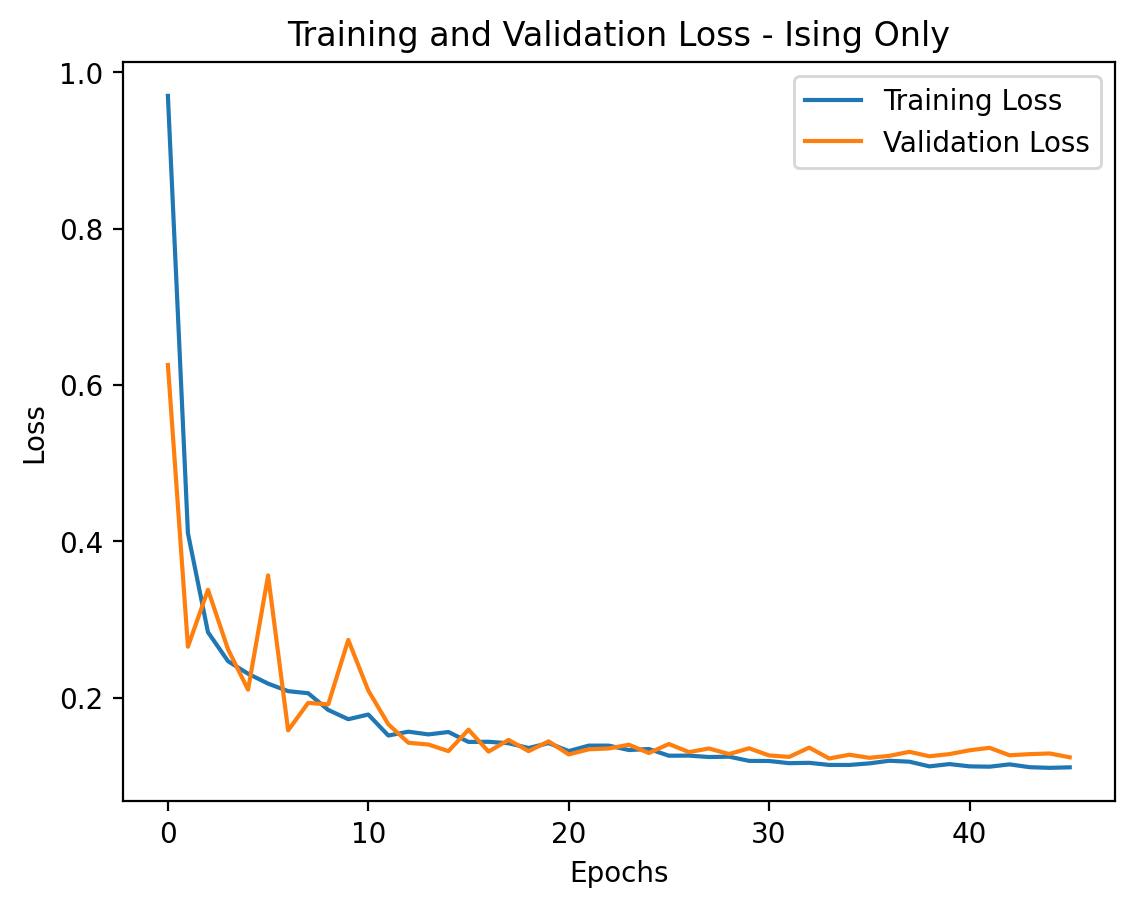}
    \hfill
    \includegraphics[width=0.48\textwidth]{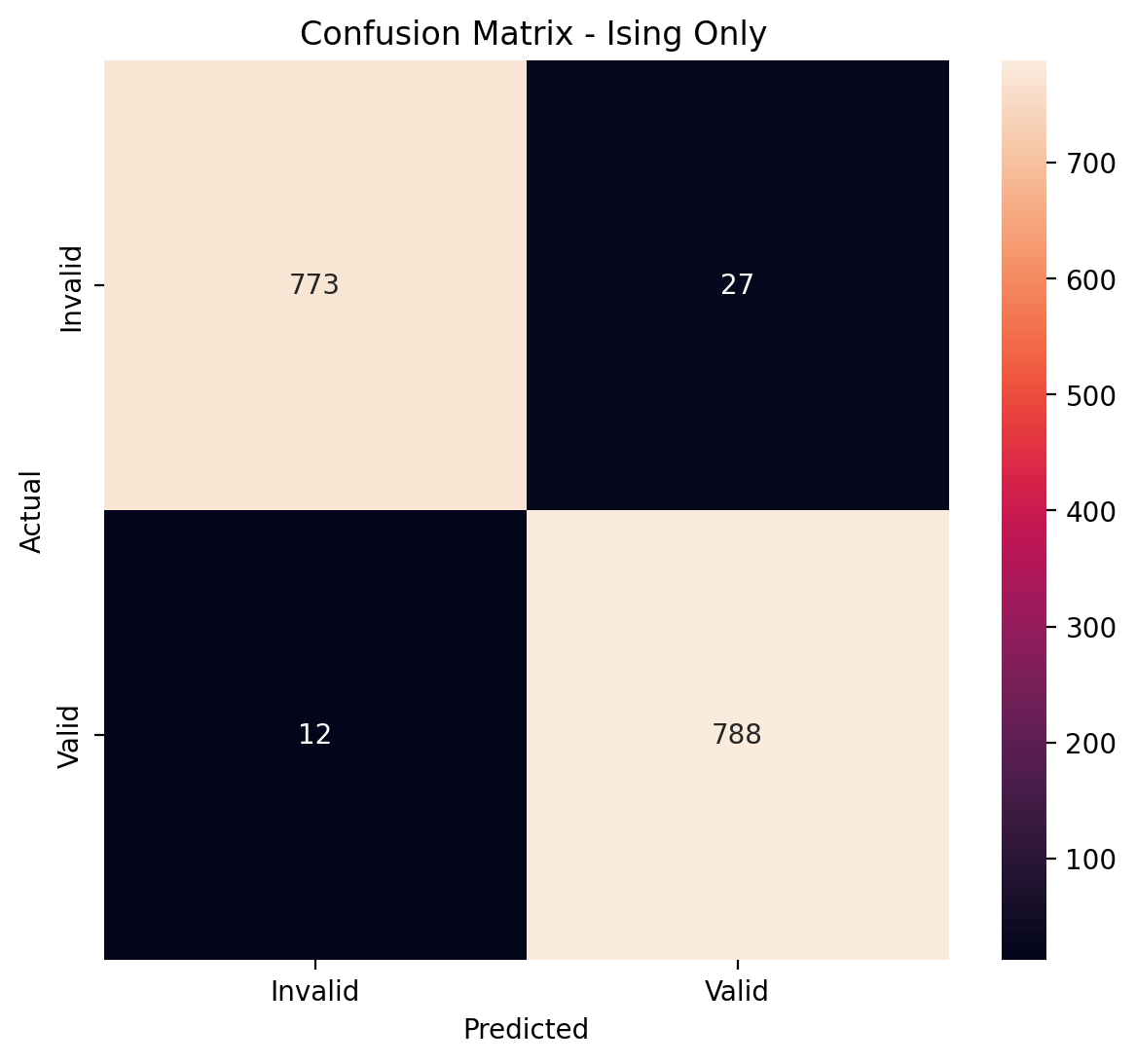}
    \caption{
    Representative training diagnostics for the Ising-based model trained with 4000 equations of state per class.
    Left: training and validation binary-cross-entropy loss as functions of epoch.
    Right: The confusion matrix.
    }
    \label{fig:ising_training_diagnostics}
\end{figure*}

\begin{figure*}[t]
    \centering
    \includegraphics[width=0.48\textwidth]{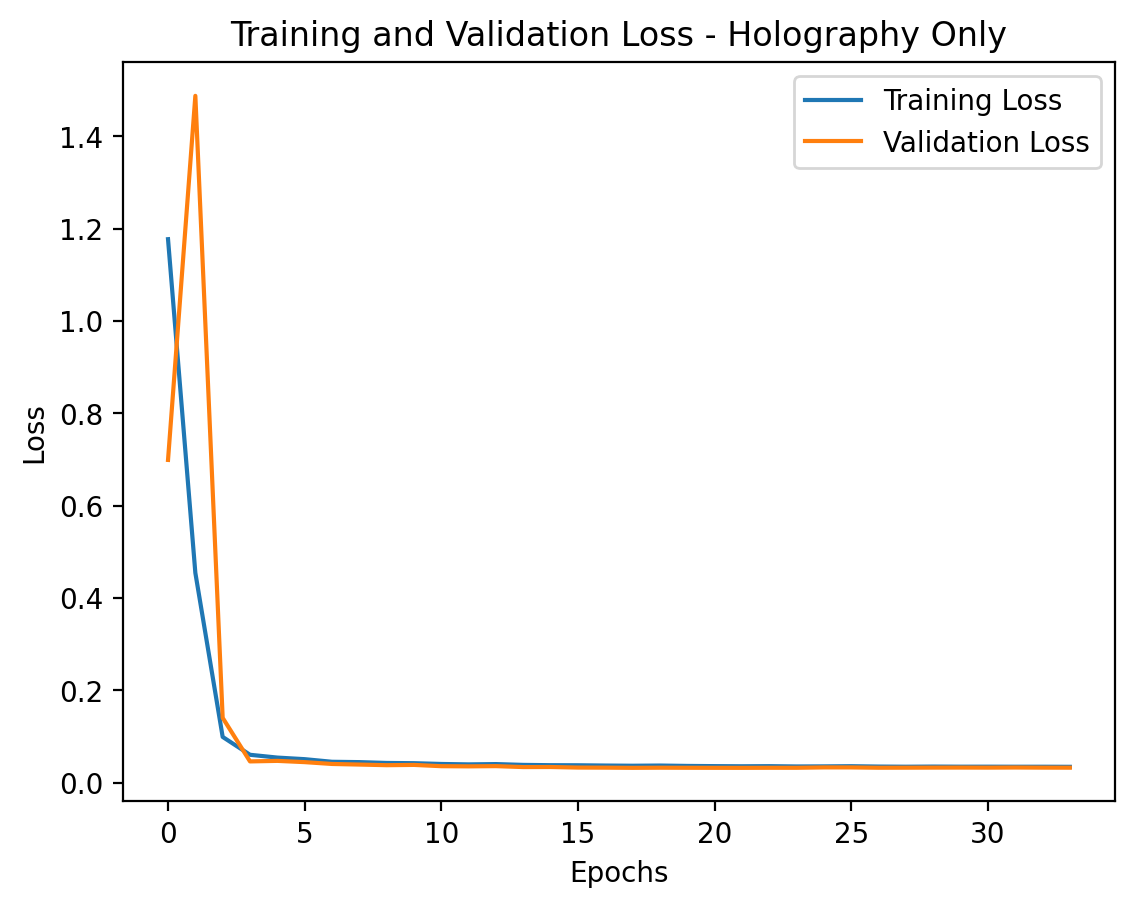}
    \hfill
    \includegraphics[width=0.48\textwidth]{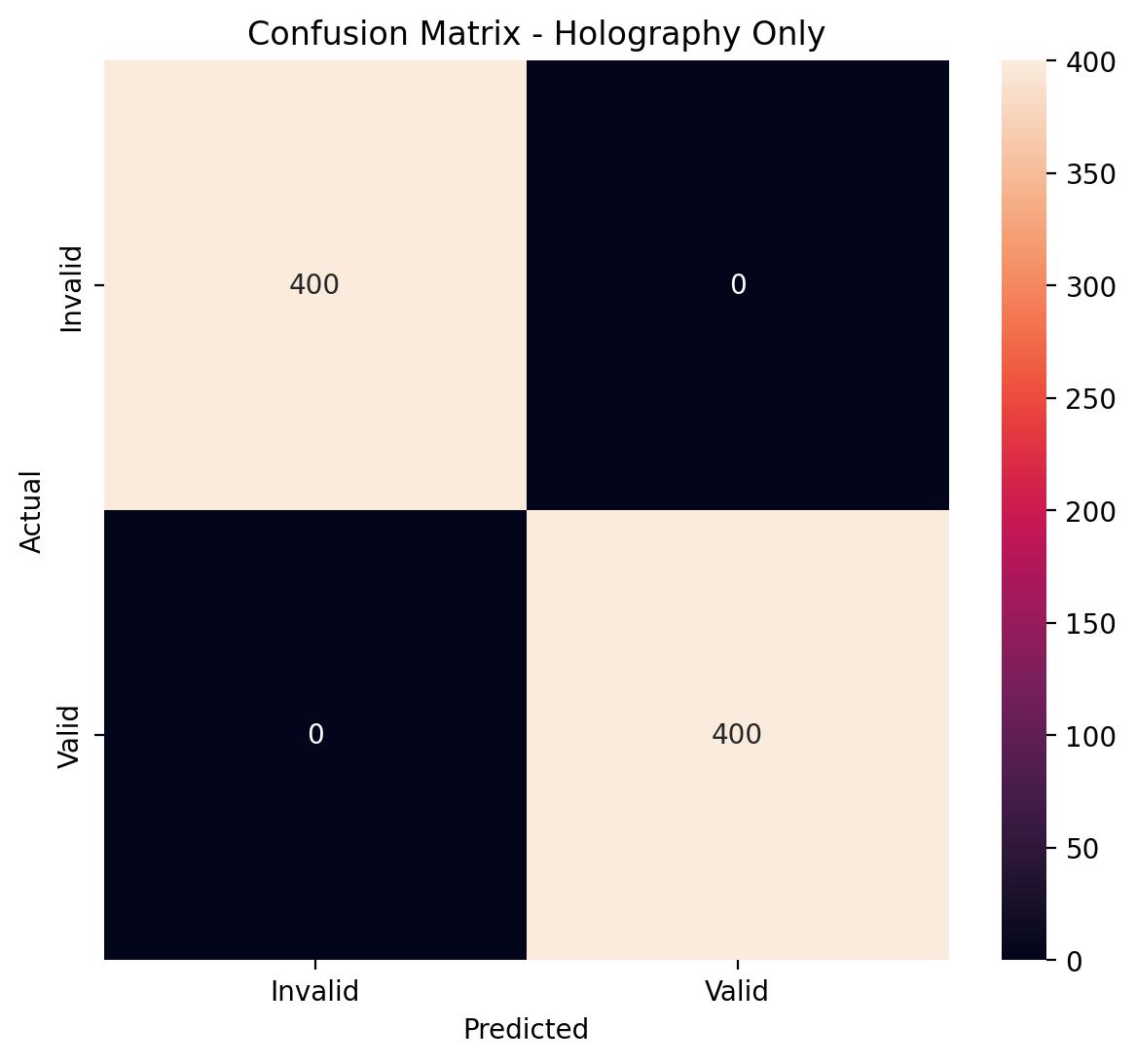}
    \caption{
    Representative training diagnostics for the holography-based model trained with 2000 equations of state per class.
    Left: training and validation binary-cross-entropy loss as functions of epoch.
    Right: The confusion matrix.
    }
    \label{fig:holography_training_diagnostics}
\end{figure*}

To further examine the learning dynamics of the classifier, representative loss curves and confusion matrices for the largest training configurations in both physical frameworks are shown in
Figs.~\ref{fig:ising_training_diagnostics} and~\ref{fig:holography_training_diagnostics}. For the Ising-based framework, the training process exhibits gradual convergence with noticeable fluctuations in the validation loss during the early training stages, suggesting a comparatively more challenging classification task between physically admissible and inadmissible EoSs. The corresponding confusion matrix nevertheless demonstrates highly balanced classification performance, with only a small number of misclassified samples. 

In contrast, the holographic framework exhibits extremely rapid convergence, with both training and validation losses stabilizing after only a few training epochs. The corresponding confusion matrix shows complete separation between the two classes, with no misclassified samples. These observations are consistent with the interpretation discussed below that, for the datasets considered here, the holographic EoSs present a simpler classification problem for the neural network than the Ising-based EoSs.

\subsection{Framework-dependent classification complexity}

The different behavior observed between the Ising-based and holographic frameworks suggests that the learnability of physical admissibility conditions depends strongly on the diversity and geometric structure of the EoS family generated by the underlying physical model.

For the Ising-based framework, both physically admissible and inadmissible EoSs are generated within the same theoretical construction through variation of the mapping parameters. This framework spans a highly diverse family of EoSs, since the mapping parameters control the location of the critical point and the structure of the transition region in the QCD phase diagram. Consequently, the resulting pressure surfaces can exhibit more diversity even within the same validity class, also valid and invalid EoSs may occupy complex and partially overlapping regions of the high-dimensional input. The classifier must therefore learn the admissibility boundary across a structurally diverse family of pressure surfaces, which is consistent with the larger number of training examples required for robust classification.

In contrast, the holographic framework considered in this work spans a less diverse family of EoSs. The underlying holographic and HRG equations of state are fixed, while different EoSs are generated by varying the parameters controlling their merging. An important feature of this construction is that the statistical-mixture prescription \cite{Yang:2026brr} produces only physically admissible EoSs over the parameter space considered: both the HRG and holographic descriptions are individually physically admissible in their respective domains, and the statistical-mixture procedure preserves this admissibility. Consequently, this framework alone cannot provide the two classes required for supervised binary classification. To construct the inadmissible class, we therefore employ an alternative switching-function prescription for merging the same HRG and holographic descriptions. Within the parameter space considered here, this construction produces inadmissible EoSs and is consequently used to generate the invalid class. Despite this distinction, both classes are generated from the same underlying HRG and holographic descriptions, with the resulting family of EoSs remaining comparatively restricted. Their pressure surfaces therefore span a more limited structural diversity than the Ising-based EoS family, consistent with the substantially smaller training set required for accurate classification.

These observations suggest that the effectiveness of machine-learning classification of admissible and inadmissible EoSs depends not only on the neural-network architecture itself, but also on the diversity of the EoS family generated by the underlying physical framework.

\subsection{Validation on Ising 2D-TExS \texorpdfstring{$(w,\rho)$}{(w,rho)} maps}

To further examine the generalization capability of the classifier within the Ising-based framework, we reconstruct physical admissibility regions in previously unseen regions of the underlying $(w,\rho)$ parameter space associated with the critical Ising mapping construction. This test provides a more stringent evaluation than the independent classification experiments discussed previously, since it allows the learned classifier to be examined over broad regions of the physical parameter space from which the EoSs are generated.

It is important to emphasize that the neural network has no access to the physical parameters $(w,\rho)$ at any stage of training or inference. As in all previous experiments, the classifier receives only the normalized pressure surface $Q(T,\mu_B)$ as input and returns a probability representing the likelihood that the EoS satisfies the physical admissibility conditions. The parameters $(w,\rho)$ are known only during the construction of the independent testing dataset.

\begin{figure*}[t]
    \centering
    \includegraphics[width=\textwidth]{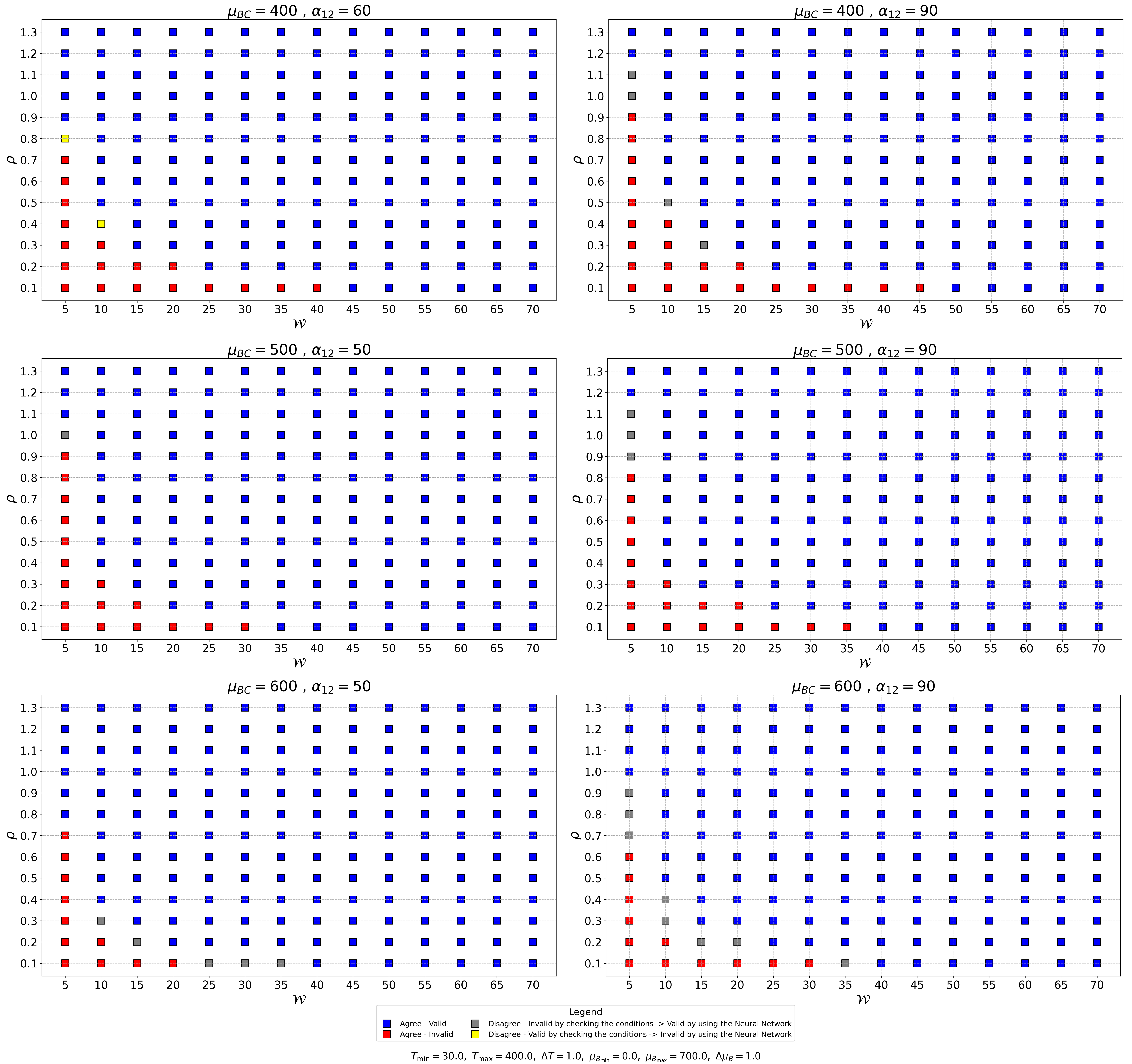}
    \caption{
    Reconstruction of physical admissibility regions in the $(w,\rho)$ parameter space for the Ising-based framework.
    Each point corresponds to an independently generated EoS.
    The maps compare the physical ground-truth label obtained from direct thermodynamic stability and causality checks with the classification predicted by the trained neural network using only the normalized pressure surface $Q(T,\mu_B)$ as input.
    Blue and red markers represent agreement between the direct physical check and the neural network prediction, with blue indicating samples classified as valid by both methods and red indicating samples classified as invalid by both. Gray and yellow markers represent disagreement: gray denotes samples that are physically invalid but predicted as valid by the neural network (false positives), whereas yellow denotes samples that are physically valid but predicted as invalid by the neural network (false negatives).
    }
    \label{fig:ising_maps}
\end{figure*}

The parameter-space maps shown in Fig.~\ref{fig:ising_maps} are therefore not direct outputs of the machine-learning model, but rather the result of a combined physical and machine-learning analysis. For each independently generated EoS, the corresponding $(w,\rho)$ coordinates are known during dataset preparation, while the physical ground-truth label is determined separately through direct numerical evaluation of the thermodynamic stability and causality conditions described in Sec.~\ref{sec:Physical Constraints}. The trained neural network then independently predicts the classification label using only the pressure surface information. The final parameter-space maps are constructed by comparing the physically determined ground-truth label with the machine-learning prediction for each EoS and assigning the comparison result to the corresponding $(w,\rho)$ coordinate.

As shown in Fig.~\ref{fig:ising_maps}, the overwhelming majority of EoSs are classified correctly throughout the explored parameter space. Large contiguous regions exhibit complete agreement between direct physical validation and machine-learning prediction, demonstrating that the classifier successfully generalizes to previously unexplored regions of the Ising parameter space despite having no direct knowledge of the physical parameters used to generate the EoSs.

A particularly important observation is that the majority of disagreement points are concentrated near the boundary separating physically valid and invalid regions. In these transition regions, the thermodynamic constraints used to define validity, such as positivity of the specific heat $C_V$, baryon susceptibility $\chi_2^B$, and the causality condition $0 \leq c_s^2 \leq 1$, are often only marginally satisfied or violated. Since these quantities are obtained through numerical differentiation of the pressure surface, small discretization effects, numerical instabilities, or finite-resolution artifacts may lead to apparent violations even when the underlying equation of state lies very close to the physical validity boundary.

This behavior suggests that the classifier is not merely reproducing the numerical labels assigned during dataset construction, but is instead learning broader geometric signatures associated with thermodynamic stability directly from the pressure surface itself. In particular, the neural network appears to smooth over sharp numerically induced boundaries and identify physically meaningful large-scale structures within the EoS surface. The concentration of disagreement points near boundary regions therefore provides further evidence that the learned representation captures physically relevant information beyond the local numerical artifacts that may affect explicit derivative-based validation procedures.

\subsection{Computational cost of EoS validation}

Beyond classification accuracy, an important practical consideration is the computational cost required to determine the physical admissibility of a candidate EoS. To quantify this aspect, we compare the execution time of the conventional physics-based validation procedure with that of the proposed machine-learning approach using pressure inputs at different stages of preprocessing. In the physics-based approach, the admissibility of each EoS is determined by explicitly evaluating the thermodynamic stability and causality conditions through the corresponding thermodynamic quantities, whereas the machine-learning approach determines the classification from the pressure surface alone.

To ensure a fair comparison, all computational pipelines are evaluated under identical computational conditions using the same GPU hardware and the same balanced collection of 2150 EoSs, consisting of 1075 physically admissible and 1075 physically inadmissible cases. The EoSs are processed sequentially, one at a time, without batching or parallelization. For the machine-learning pipelines, model loading and initialization are performed before timing, such that the measured execution time includes only the processing required to transform the specified pressure representation into a validity classification.

Table~\ref{tab:timing_comparison} summarizes the average execution time
required to determine the physical admissibility of a single equation of
state using the conventional physics-based validation procedure and the
different machine-learning validation pipelines. Four machine-learning
input representations are considered: the raw pressure stored as a
\texttt{.dat} file, the pressure surface stored as a two-dimensional
\texttt{.npy} array, the normalized pressure surface
$Q=P/T^4$ stored as a two-dimensional \texttt{.npy} array, and the
standardized pressure surface $\widetilde{Q}$ stored as a two-dimensional
\texttt{.npy} array. For each computational pipeline, the reported values
represent the mean processing time per EoS and the corresponding standard
deviation over ten runs.

\begin{table}[t]
\centering
\caption{
Average execution time required to determine the physical admissibility
of a single equation of state. The reported values are the mean and
standard deviation over ten runs.
}
\label{tab:timing_comparison}
\begin{tabular}{lc}
\hline\hline
Method & Average time per EoS (ms) \\
\hline

Physics-based validation
    & $113.62 \pm 0.33$ \\

ML (Raw Pressure File, \texttt{.dat})
    & $59.64 \pm 5.45$ \\

ML (2D Array Pressure, \texttt{.npy})
    & $6.33 \pm 0.14$ \\

ML (2D Array $Q=P/T^4$, \texttt{.npy})
    & $6.26 \pm 0.25$ \\

ML (2D Array $\widetilde{Q}$, \texttt{.npy})
    & $\mathbf{5.59 \pm 0.15}$ \\
\hline\hline
\end{tabular}
\end{table}

The results presented in Table~\ref{tab:timing_comparison} demonstrate
that the proposed machine-learning approach substantially reduces the
computational cost of EoS validation. The conventional physics-based
procedure requires an average of $113.62 \pm 0.33$ ms per EoS, whereas
the ML pipeline starting directly from a raw pressure \texttt{.dat} file
requires $59.64 \pm 5.45$ ms per EoS, corresponding to a speedup of
approximately $1.9$. A much larger reduction is obtained when the
pressure surface is already available as a two-dimensional array. Starting
from $P(T,\mu_B)$, $Q(T,\mu_B)=P(T,\mu_B)/T^4$, and the standardized
pressure surface $\widetilde{Q}(T,\mu_B)$, the average processing times
are $6.33 \pm 0.14$, $6.26 \pm 0.25$, and $5.59 \pm 0.15$ ms per EoS,
respectively. The fastest pipeline, which takes $\widetilde{Q}$ directly
as input, is therefore approximately $20.3$ times faster than the direct
physics-based validation performed under the same GPU hardware conditions.

The big difference between the raw \texttt{.dat} and
two-dimensional \texttt{.npy} ML pipelines shows that a large fraction
of the computational cost when starting from the raw pressure file is
associated with reading and preparing the input rather than with the
neural-network evaluation itself. Once the pressure surface is available
as a two-dimensional array, the processing time decreases from
$59.64 \pm 5.45$ ms to approximately $6$ ms per EoS. In contrast, the
small differences among the $P$, $Q$, and $\widetilde{Q}$ array-based
pipelines indicate that the transformations required to construct
$Q=P/T^4$ and subsequently standardize it introduce comparatively little
additional computational overhead. The reported execution times therefore
characterize the complete validation pipeline starting from each specified
input representation, rather than the neural-network inference time alone.

This computational advantage becomes particularly relevant in large
parameter-space explorations, where thousands or even millions of
candidate EoSs may need to be evaluated. In such applications, the
proposed classifier can serve as a fast screening tool for large-scale
EoS generation pipelines, rapidly identifying candidate EoSs according
to their physical admissibility while reserving explicit thermodynamic
stability and causality calculations for cases requiring detailed
physics-based verification.

\section{Conclusions}

In this work, we investigated whether the physical admissibility of EoSs can be inferred directly from the geometry of the pressure surface using deep learning. Rather than explicitly evaluating the thermodynamic quantities traditionally employed to verify stability and causality, we trained a convolutional neural network using only the normalized pressure surface, $Q(T,\mu_B)=P(T,\mu_B)/T^{4}$, while the physical labels were assigned independently through direct evaluation of the thermodynamic stability and causality conditions. The network therefore receives no information about the specific heat, baryon-number susceptibility, speed of sound, higher-order thermodynamic derivatives, or the internal parameters of the theoretical framework used to generate the EoS.

Our results demonstrate that the normalized pressure surface alone contains sufficient geometric information for reliable identification of physically admissible EoSs. When trained on EoSs generated within the Ising-based framework, the classifier achieves an independent testing accuracy of 97.65\% on previously unseen EoSs. Repeating the entire training and evaluation procedure using an independent holography framework (merged with hadron resonance gas model at low temperature) yields perfect classification on the corresponding independent test set. Together, these results show that the proposed methodology is not tied to a particular EoS construction, but instead learns geometric signatures associated with thermodynamic stability and causality that are encoded in the pressure surface itself.

An important outcome of this work is that it establishes a proof of concept for a new strategy for EoS validation. The classifier is never trained on the thermodynamic observables used to define physical admissibility, such as the specific heat, baryon-number susceptibility, or the speed of sound. Instead, these quantities are used only once to generate the ground-truth labels during dataset construction. After training, the pretrained network requires only the normalized pressure surface as input to determine whether an EoS is physically admissible. This demonstrates that explicit evaluation of higher-order thermodynamic quantities is not required during inference, opening the possibility of replacing expensive derivative-based validation with direct classification based solely on the geometry of the pressure surface.

Beyond its classification accuracy, the proposed approach also provides a significant computational advantage. Benchmark calculations performed under identical computational conditions show that the pretrained classifier substantially reduces the computational cost of EoS validation compared with conventional physics-based verification through explicit stability and causality checks. Such acceleration becomes increasingly valuable in large parameter-space explorations, where thousands or millions of candidate EoSs may need to be screened before more detailed thermodynamic analyses are performed.

Although the present work employs a specific convolutional neural network architecture and training procedure, the methodology itself is considerably more general. Any research group developing EoSs using different microscopic theories, effective models, or numerical frameworks can generate their own labeled datasets using the same physical admissibility criteria, train the same neural-network architecture, and obtain a framework-specific classifier tailored to their own EoS family. In this sense, the proposed approach provides a general machine-learning workflow for rapid EoS validation rather than a model restricted to the particular datasets considered in this study.

There are several natural directions for future work. The present classifier can be further optimized toward either higher classification accuracy and lower computational cost through improvements in network architecture, data representation, or inference strategies. It will also be interesting to investigate additional EoS frameworks, larger parameter spaces, uncertainty quantification, and extensions beyond binary classification. More broadly, the methodology introduced here demonstrates how machine learning can complement traditional thermodynamic calculations by learning physically meaningful geometric representations directly from the pressure surface. As part of future developments, the classifier developed in this work is planned to be integrated into the MUSES Calculation Engine \cite{Jahan:2026hvs}, where it will serve as a fast screening module for identifying physically admissible EoSs before more computationally expensive physics-based analyses are carried out.

\begin{acknowledgments}
This material is based upon work supported by the National Science Foundation under grants No. PHY-2208724, PHY-2116686, PHY-2514763, PHY-2621752 and PHY-2623480, and within the framework of the MUSES collaboration, under Grant No. OAC-2103680. This material is also based upon work supported by the U.S. Department of Energy, Office of Science, Office of Nuclear Physics, under Award Number DE-SC0022023, as well as by the National Aeronautics and Space Agency (NASA) under Award Number 80NSSC24K0767.
\end{acknowledgments}

\bibliographystyle{unsrt}
\bibliography{bibfile}

\end{document}